\documentclass[%
 reprint,
superscriptaddress,
 amsmath,amssymb,
 aps,
 prl,
]{revtex4-2}

\usepackage{bbold}
\usepackage{graphicx}
\usepackage{dcolumn}
\usepackage{bm}
\usepackage{xcolor}
\DeclareMathOperator{\sgn}{sgn}

\begin{document}

\title{Ising machine by dimensional collapse of nonlinear polarization oscillators}

\author{Salvatore Chiavazzo}
\affiliation{Institute for Complex Systems, National Research Council (ISC-CNR), 00185 Rome, Italy} 

\author{Marcello Calvanese Strinati}
\affiliation{Enrico Fermi Research Center (CREF), 00184 Rome, Italy}

\author{Claudio Conti}
\affiliation{Department of Physics, Sapienza University of Rome, 00185 Rome, Italy}

\author{Davide Pierangeli}
\email{davide.pierangeli@roma1.infn.it}
\affiliation{Institute for Complex Systems, National Research Council (ISC-CNR), 00185 Rome, Italy} 
\affiliation{Department of Physics, Sapienza University of Rome, 00185 Rome, Italy}

\begin{abstract}

Ising machines show promise as ultrafast hardware for optimizations encoded in Ising Hamiltonians but fall short in terms of success rate and performance scaling. 
Here, we propose a novel Ising machine that exploits the three-dimensional nature of nonlinear polarization oscillators to counteract these limitations.
Based on the evolution of the optical polarization in third-order nonlinear media, the high-dimensional machine reaches the Ising ground state by the mechanism of \emph{dimensional collapse}: the dynamics on the Poincaré sphere undergoes a self-induced collapse into polarization fixed points mapping Ising spins.
The photonic setup consists of polarization-modulated pulses in a $\chi^{(3)}$ crystal
subject to iterative feedback. 
We numerically demonstrate that its high-dimensional operation leads to an enhanced success probability on benchmark graphs and an exponential improvement in performance scaling with respect to coherent Ising machines.
The proposed polarization Ising machine paves the way for a new class of 
Ising solvers with enhanced computing capabilities.
\end{abstract}

\maketitle

Physical systems that implement spin Hamiltonians are attracting vast attention as unconventional computing paradigms
to overcome the time and energy limitations of digital hardware in combinatorial optimization and machine learning.
Ising machines (IMs)~\cite{McMahon2022} can impact a myriad of applications 
by accelerating the ground state (GS) search of the Ising model encoding the NP-hard problem~\cite{Lucas2014}.
Realizations range from classical and quantum annealers made by 
optical devices~\cite{Pierangeli2019, Soljacic2020, Pierangeli2020, Pierangeli2021, Leonetti2021, Suzuki2023, Ruan2023, Yan2024, Veraldi2025}, superconducting circuits~\cite{Boixo2014}, magnetic junctions~\cite{Datta2019}
and memristors~\cite{Strachan2020},
to dynamical-system solvers made by networks of electronic~\cite{Wang2019, Herzog2019, Peer2019, Cassella2024, Eichler2024, Datta2021, English2022, Kim2023}, acoustic~\cite{Akerman2023}, and optical oscillators~\cite{Marandi2014, Inagaki2016},
lasers~\cite{Babaeian2019, Davidson2020} and polaritons~\cite{Berloff2017}, 
whose computing principles have led to novel heuristics~\cite{Goto2019, Carmes2020, Kalinin2018, Leleu2021}.
Photonic IMs~\cite{Danner2024} such as coherent IMs (CIMs) based on degenerate optical parametric oscillators~\cite{Yamamoto2013, McMahon2016, Takesue2016, Hamerly2019, Takesue2021, Gaeta2020, Strinati2021_2}
and optoelectronic oscillators~\cite{Vandersande2019, MingLi2022, Guangzhou2024} 
are especially promising for achieving ultrafast and large-scale solutions
by virtue of optical bandwidth and parallelism. 

A critical limitation of IMs is their trapping in local minima 
due to the approximate embedding of the Ising Hamiltonian within the dynamics of the analog platform.
This leads to sub-optimal solutions and restricts the functioning to narrow conditions,
which results in a reduced success rate and poor performance scaling.
Intensive research is aimed at improving IMs 
performance~\cite{Leleu2019,Strinati2021_1,Bohm2021,Bifurcation2023}.
Recent works have proposed the use of multidimensional spins~\cite{Strinati2022, Strinati2024, Berloff2024} to leverage the system's additional dimensions as an escaping mechanism leading to the GS~\cite{Strinati2024}.
These approaches stimulate the idea of building superior IMs by using high-dimensional ($D\geq3$) instead of low-dimensional oscillators.
However, a system of high-dimensional oscillators that map Ising spins has yet to be identified.

In this Letter, we propose a high-dimensional IM based on a novel type of optical oscillator: nonlinear polarization oscillators (NPOs) in third-order nonlinear media.
We exploit NPOs as three-dimensional computing units represented by the Poincaré sphere.
Unlike current IMs that encode spins in phase or amplitude,
our approach maps spherical spins in the polarization of the optical field.
We design the polarization IM (PIM) as a network of coupled NPOs.
The setup consists of time-multiplexed pulses modulated in polarization that evolve iteratively in a $\chi^{(3)}$ crystal subject to measurement and feedback. 
The PIM operates through a transition on the Poincaré sphere
with NPOs that spontaneously collapse into a binary steady-state 
configuration mapping the Ising GS.
We refer to this novel high-dimensional computing mechanism as \emph{dimensional collapse}.
Extensive simulations demonstrate enhanced success probability and exponentially 
improved scaling of performance metrics on benchmark Max-Cut 
and complete graphs of hundreds of spins. 
Exploiting the extra dimensions to escape from local minima, 
the PIM can solve problems for which CIMs hardly converge to the GS.
The simplicity of the PIM setup and the broad applicability of its operating principle
suggest a promising route to realize high-dimensional IMs.

NPOs describe the dynamics of the optical polarization in nonlinear media~\cite{Zheludev2000}. 
We consider light at frequency~$\omega$ propagating in a lossless crystal
with anisotropic susceptibility tensor $\chi^{(3)}$ and no birefringence [Fig. 1(a)]. 
The $z$-evolution of the Stokes vector $\mathbf{S}=(S_0, S_1, S_2, S_3)$
admits the motion invariants $S_{\rm 1}^2 + S_{\rm 2}^2/\eta = r^2 $ and $ S_{\rm 1}^2 + S_{\rm 2}^2 + S_{\rm 3}^2 = S_{\rm 0}^2$, being $\dot S_{\rm 0}=dS_{\rm 0}/dz=0$ where $S_{\rm 0}$ is the optical intensity, which results in the NPO equation~\cite{Wabnitz1986} 
\begin{equation}    
\ddot S_{\rm i} + \alpha_{\rm i} S_{\rm i} + \beta_{\rm i} S_{\rm i}^3 = 0 ,
\end{equation}
with ${\rm i}=1,2,3$. The oscillator coefficients read as
$\alpha_{\rm 1}=\chi^2 \eta \left[ S_{\rm 0}^2 + (1 - 2 \eta ) r^2 \right]$ 
and $\beta_{\rm 1}= 2 \chi^2 \eta (\eta - 1)$,
and similar for ${\rm i}=2,3$,
where $\chi$ is the nonlinear coefficient and $\eta$ the anisotropy factor,
as detailed in Supplemental Material (SM)~\cite{suppli} .
The NPO of Eq.~(1) has four stable fixed points at $S_{\rm 1}=\pm S_{\rm 0}$ and $S_{\rm 3}=\pm S_{\rm 0}$, 
and two saddle points at $S_{\rm 2}=\pm S_{\rm 0}$.
Trajectories on the Poincaré sphere for an input vector in proximity of these points are shown in Fig.~1(b)-(c) by numerical integration of Eq.~(1) with $S_{\rm 0}=5$ GW/cm$^2$, $\chi=1$ cm/GW, $\eta=1.1$ (e.g. a KTa$_{x}$Nb$_{1-x}$O$_3$ crystal at wavelength 
$\lambda=633$~nm).
The instability with respect to the $S_{\rm 2}$ axis result into polarization symmetry breaking~\cite{Zheludev1989}, with the emergence of different orbits depending on input polarization fluctuations [Fig.~1(c)]. 
Orbits around the stable points (centres) [Fig.~1(b)] 
offer robust ways to encode binary information in the NPO.

\begin{figure}[t!]
\centering
\includegraphics[width=0.95\columnwidth]{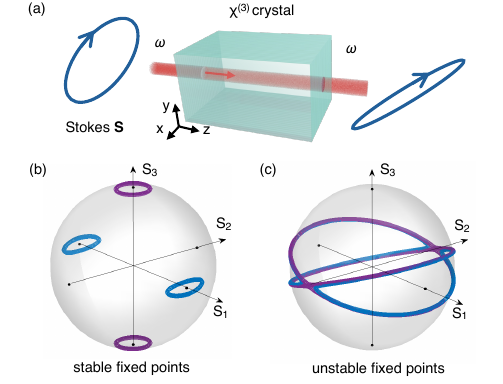} 
\vspace*{-0.1cm}
\caption{{Nonlinear polarization oscillator (NPO).} 
(a) Polarization evolution by nonlinear propagation in a third-order anisotropic crystal. 
Trajectories on the Poincaré sphere by solving Eq.~(1)
for input Stokes vectors near of the (b) stable and (c) unstable fixed points. 
}
\vspace{-0.2cm}
\label{Figure1}
\end{figure}

To build an IM based on NPOs, we introduce their iterative dynamics.
We design an optoelectronic feedback system in which $N$ optical pulses of variable polarization and intensity
iteratively propagate in the $\chi^{(3)}$ crystal of length $L$.
The Stokes vector of the $j$-th pulse ${\bf S}_j$ is modulated 
according to a measurement and feedback scheme.
The setup, illustrated in Fig.~2(a), includes a polarization modulator, i.e. electro-optic phase and amplitude modulators in series, and a full-Stokes polarimeter. 
A digital processor such as a field programmable gate array
processes the measurements and programs the modulators, coupling the NPOs.
Remarkably, as we only modulate the laser at each iteration,
the scheme does not require pulse synchronization and phase coherence.
We modulate the $j$-th input Stokes vector following the nonlinear iterative map
\begin{equation}
\begin{split}
& S_{\rm 1 \it j}^{k+1}[0]= aS_{\rm 1 \it j}^{k}[L] + bf^{k}_j ,\\
& S_{\rm 2 \it j}^{k+1}[0]= aS_{\rm 2 \it j}^{k}[L] ,\\
& S_{\rm 3 \it j}^{k+1}[0]= cS_{\rm 3 \it j}^{k+1}[L] ,\\
\end{split}
\end{equation}
where $S_{\rm i \it j}^{k}[0]$ ($S_{\rm i \it j}^{k}[L]$) is the $\rm i$-th Stokes component at the $k$-iteration
at the crystal input (output), $a<1$ and $c<1$ are the linear and circular loss parameters, 
$f^{k}_j$ is the feedback signal, $b$ is the feedback strength,
and $\left(S_{\rm 0 \it j}^{k+1}[0]\right)^2= \left(S_{\rm 1 \it j}^{k+1}[0]\right)^2 + \left(S_{\rm 2 \it j}^{k+1}[0]\right)^2  + \left(S_{\rm 3 \it j}^{k+1}[0]\right)^2$.
To study a large network of NPOs evolving by Eq.~(2),
we discretize Eq.~(1) and evaluate the output $S_{\rm i \it j}^{k}[L]$ as
\begin{equation}
S_{\rm i \it j}[L]= S_{\rm i \it j}[0] + L \dot S_{\rm i \it j}[0] - L^2 \alpha_{\rm i \it j}S_{\rm i \it j}[0] - L^2 \beta_{\rm i \it j} S^3_{\rm i \it j}[0] .
\end{equation}
The approximation accurately models the NPO for $L\ll 2\pi/\sqrt{\alpha_{\rm i}}$ (see SM~\cite{suppli}) and allows us to efficiently simulate a network of $N$ NPOs by avoiding the integration of $N$ Eqs.~(1) at each iteration. 

\begin{figure}[t!] 
\centering
\includegraphics[width=1.0\columnwidth]{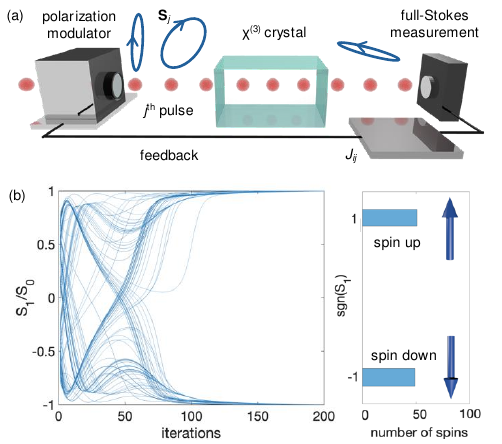} 
\caption{{Polarization Ising machine (PIM).}
(a) Scheme of the photonic setup.
The Stokes vector of a train of pulses is modulated at each iteration
according to polarization measurements and feedback.
(b) Iterative dynamics of $100$ uncoupled NPOs
for $\chi=0.3$ cm/GW, $\eta=1.1$, $L=1.2$ cm, $a=0.95$, $c=0.98$, $b=0.01$. 
Each NPO collapses to the stable fixed points $S_{1}/S_{0}=\pm 1$, which map an Ising spin.
Spin up and down have equal probability due to randomness in $S^0_{\rm 1}$ and $S^0_{\rm 3}$.
}
\vspace{-0.2cm}
\label{Figure2}
\end{figure}

\begin{figure*}[t!]
\centering
\hspace*{-0.3cm}
\includegraphics[width=2.1\columnwidth]{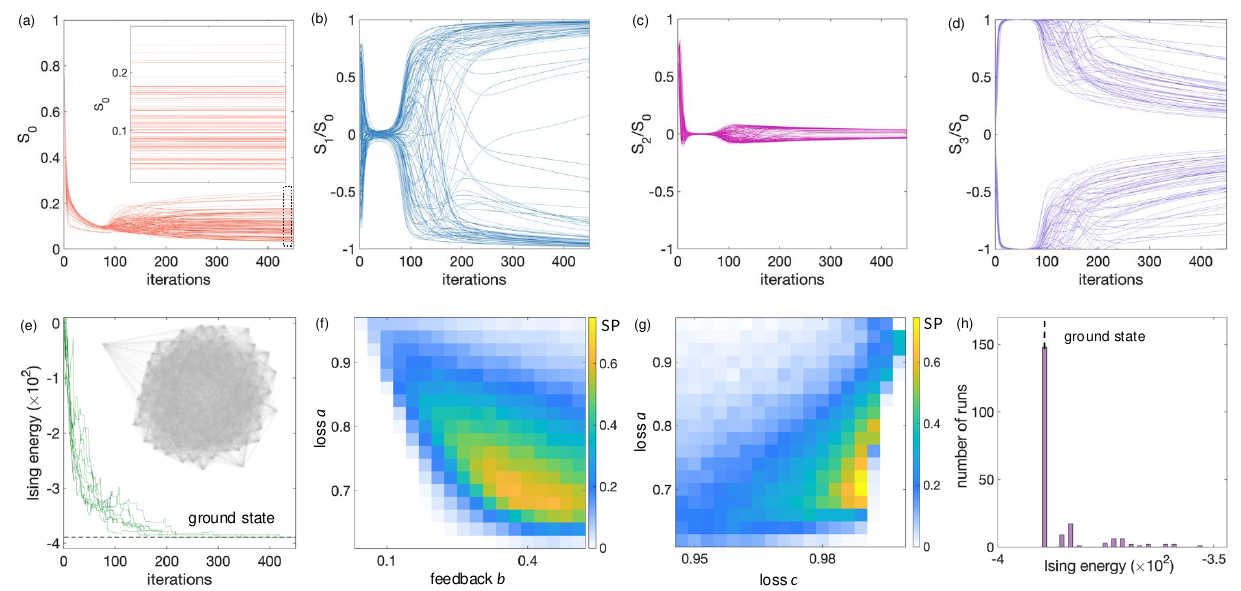} 
\vspace*{-0.5cm}
\caption{{PIM high-dimensional dynamics on a 100-spin Max-Cut problem.}
Evolution of the (a) intensity $S_{0}$ (dotted region zoomed in the inset)
and normalized Stokes parameters (b) $S_{\rm 1}/S_{0}$, (c) $S_{\rm 2}/S_{0}$, and (d) $S_{\rm 3}/S_{0}$ 
when solving the $g05_{100.2}$ benchmark graph (inset). (e) Ising energy $H$ for various runs with initial conditions varied randomly. The dashed line indicates the GS energy.
(f) Success probability SP by scanning the parameters $a$, $b$ and (g) $a$, $c$.
(h) Distribution of $H$ at the steady state over 200 runs. Results are for $a=0.70$, $b=0.40$, $c=0.99$, 
$\chi=0.3$~cm/GW, $\eta=1$, $L=1.2$~cm, $S^0_{\rm 2}/S^0_{\rm 0}=0.66$, $S_0^0= 1$~GW/cm$^2$.
}
\vspace{-0.2cm}
\label{Figure3}
\end{figure*}

The PIM is modeled by Eqs.~(2)-(3).
For $b=0$, the only fixed point is the null state $\mathbf{S}_{\rm null}=(0,0,0,0)$.
The optical field exponentially decays to zero for any initial polarization.
Non-trivial fixed points emerge as we turn on ($0<b<1$) the feedback term
\begin{equation}
f^{k}_j= \frac{1}{N}\sum_{i}^N J_{ij} S_{\rm 1 \it i}^{k}[L] /S_{\rm 0 \it i}^{k}[L] ,
\end{equation}
where $J_{ij}$ is the coupling matrix.
The feedback acts as a nonlinear gain for the $S_{\rm 1}$ component.
It breaks the intrinsic symmetry of the NPO stable points.
Due to its dependence on the inverse of $S_{0}$, 
the feedback amplitude grows as the intensity decreases
during the dynamics. The effect counterbalances the field decay.
As the balance between gain and loss occurs,
the field collapses to a stable fixed point with $S_{\rm 1}/S_{0}=\pm 1$.
To illustrate the phenomenon, 
Fig.~2(b) shows the iterative dynamics of $N=100$ uncoupled NPOs 
(self-interaction $J_{ij}= \mathbb{1}$) 
with initial ($k=0$) intensity $S_0^0= 1$~GW/cm$^2$, $S^0_{\rm 2}/S^0_{\rm 0}=0.66$, 
and random initial components $S^0_{\rm 3}$ and $S^0_{\rm 1}$.
We observe convergence to the steady state ${\bf S^*}=(S^*_{\rm 0},\pm S^*_{\rm 0}, 0, 0)$, 
with $S^*_{0}=b/\lvert a-1\lvert$.
The polarization of the $j-$th NPO at the steady state maps an Ising spin $\sigma_j$ defined by
\begin{equation}
\sigma_j = \sgn(  S_{\rm 1 \it j} /S_{\rm 0 \it j}) .
\end{equation}
The histogram in Fig.~2(b) indicates that spin up and down have equal probability,
confirming that free NPOs subject to self-feedback behave as independent Ising spins 
at the steady state.
In SM~\cite{suppli}, we prove the stability (attraction) of the fixed points $S^*_{\rm 1 \it j}=\pm S^*_0/2$ for two coupled NPOs. 
For arbitrary $J_{ij}$, collapse towards a binary configuration with $S_{1j}/S_{0j}=\pm 1$ occurs in a broad range of parameters.

To demonstrate the PIM finds the Ising GS, 
we run unweighted Max-Cut benchmark problems $J_{ij}$ from the BiqMac library~\cite{BiqMac}, which have known GS energy.
The dynamics of the Stokes components during a single run on a $100$-spin graph 
with $50\%$ edge probability ($g05_{100.2}$) is reported in Fig.~3(a)-(d) for $a=0.70$, $b=0.40$, $c=0.99$.
Figure~3(a) shows how $S_0$ decays until a critical point,
 which occurs after nearly $80$ iterations,
and then stabilizes to a steady state where each NPO has a different intensity.
This intensity heterogeneity is not detrimental due to the polarization encoding of the spin.
The $S_{\rm 1}$ components shrink to zero before undergoing a transition into the fixed points $S_{1}/S_{0}=\pm 1$ [Fig.~3(a)]. 
On $S_{3}$, we observe an initial locking to states with $S_{3}/S_{0}=\pm 1$
followed by a transition towards zero [Fig.~3(d)]. 
$S_{2}$ remains small but its sign plays a role at the critical point [Fig.~3(c)].
This high-dimensional dynamics can lead to the GS of the corresponding Ising Hamiltonian $H=-\sum_{ij} J_{ij}\sigma_i\sigma_j$. 
We map the polarization state to the spin configuration by Eq.~(5) 
and evaluate $H$ at each iteration.
The convergence of $H$ to the GS is shown in Fig.~3(e) for different successful runs.
The probability of finding the GS at the steady state defines the success probability (SP).
The SP strongly depends on the machine parameters.
In Fig.~3(f), we report the SP, averaged over 10 repetitions of 200 runs,
when scanning $a$ and $b$.
We find a broad region of the parameter space where the PIM achieves a high SP.
The SP grows by decreasing $a$ (larger linear losses),
which means that a fast initial decay of $S_1$ helps the GS search.
Increasing $b$ accelerates the convergence to the GS (see SM~\cite{suppli}).
A key role in the effectiveness is played by the circular loss~$c$.
The map in Fig.~3(g) shows a narrow region of optimal operation.
A slight decrease of $c$ causes a SP drop.
We attribute the effect to the initial growth of $S_{3}/S_{0}$.
When decreasing $c$, the states $S_{3}/S_{0}=\pm 1$ are not reached or last only a few iterations.
This evidence indicates the high-dimensional mechanism leading to the GS:
the jump to $S_{3}/S_{0}=\pm 1$ followed by the transition from these states to the fixed points. The whole process corresponds to a self-induced dynamical collapse 
from spherical to Ising spins.
After the transition point, the NPOs network sets in the configuration that minimizes the power dissipation rate according to the principle of minimum loss~\cite{Yablonovitch2020}.
Since the PIM dissipation rate maps to the Ising Hamiltonian as detailed in SM~\cite{suppli}, the GS can be found at the steady state.

The PIM achieves a very high SP on benchmark problems. 
On the $g05_{100.2}$ graph, 
tuning the parameters we find a maximum SP$_{\rm max}=0.78\pm0.03$. 
We remark that SP is evaluated only by using the outcome of the dynamics 
(without considering any transient GS) as this condition is relevant in hardware.
The final energy distribution over $200$ runs is shown in Fig.~3(f).
A SP considerably larger than simulated CIMs~\cite{VanDerSande2025}
is achieved also for other BiqMac graphs (see SM~\cite{suppli}).
Remarkably, we find a finite SP even for problems classified as `Ising hard' for IMs 
(e.g. $g05_{100.3}$). For these problems, CIMs always converge to a local minimum
in absence of noise as the GS is not connected to the bifurcation branch~\cite{VanDerSande2024}. 
In the PIM, these GSs become accessible through the extra dimensions.
The NPOs can be initialized in each run in different polarizations 
and hence their collapse has multiple outcomes.
Results varying the initial Stokes components (see SM~\cite{suppli})
indicate that the GS search can benefit from a larger input space.
Moreover, we observe that a minimum length $L$ and nonlinearity $\chi$ are required to reach the GS, which confirms that NPO propagation is essential for the PIM computing capability. When varying $\eta$, SP$_{\rm max}$ is found for $\eta=1$ (isotropic crystal).

\begin{figure}[t!]
\centering
\hspace*{-0.4cm}
\includegraphics[width=1.0\columnwidth]{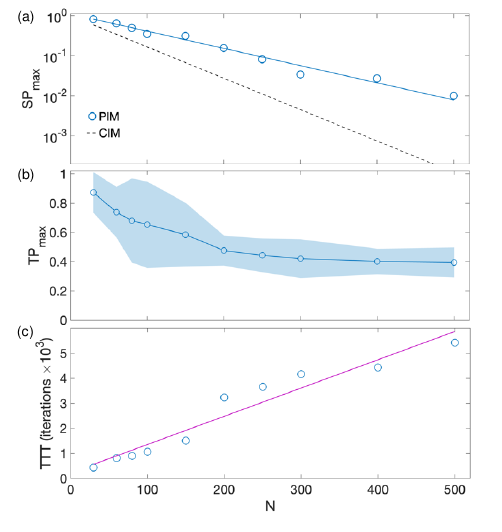} 
\vspace*{-0.4cm}
\caption{Scaling of the PIM computational performance. (a)~SP$_{\rm max}$ on K graphs of size $N$. 
The results (dots) are fitted by an exponential function (blue line) with decay rate $\gamma_{\rm PIM}=9 \pm 1 \times 10^{-3}$. The finite-size scaling of SP$_{\rm max}$ for CIMs from~\cite{Strinati2024} (dotted line) 
is reported to show the PIM improvement in performance.
(b) Accuracy in providing low-energy solutions TP$_{\rm max}$. The shaded area is the interquartile range.
(c) The time-to-target TTT (dots) grows as a linear function of $N$ (magenta line).
}
\vspace{-0.2cm}
\label{Figure4}
\end{figure}

To study the scaling of the PIM computational performance, 
we run complete graphs with random $J_{ij}=\pm 1$ (K graphs)
 of increasing size up to $N=500$.
The SP is evaluated using the GS found via simulated annealing.
Figure~4(a) reports SP$_{\max}$ when scanning $a$, $b$, and $c$, averaged on 20 random graphs for each $N$.
SP$_{\max}$ shows an exponential decay well fitted by the exponential function 
$p_s(N)\propto e^{-\gamma_{\rm PIM} N}$ with $\gamma_{\rm PIM}=9 \pm 1 \times 10^{-3}$.
We compare this decay with the exponential decay of CIMs on K graphs,
for which a finite-size scaling analysis (up to $N=100$) gives $\gamma_{\rm CIM}\simeq18 \times 10^{-3}$~\cite{Strinati2024}.
The PIM shows a decrease of SP$_{\max}$ that is exponentially slower,
with an enhancement given by the decay rate ratio $\gamma_{\rm CIM}/\gamma_{\rm PIM}\approx2$.
We ascribe the exponential improvement in performance over CIMs 
to the high-dimensional operation of the PIM.
A similar scaling advantage has been recently reported in the hyperspin machine with dimensional annealing~\cite{Strinati2024}, where the Ising model is embedded in a higher dimension by using parametric oscillators behaving as circular spins.
However, while dimensional annealing adiabatically changes selected couplings 
to effectively reduce the dimensionality during the evolution,
the PIM dimensional collapse from 3D to 1D spins is induced by the feedback mechanism 
and occurs spontaneously without an external annealing technique.

We complete the performance analysis with additional figures of merit.
To quantify the mean accuracy, we use the target probability TP, 
defined as the probability of finding a steady state of energy within $1\%$ 
of the GS energy.
The scaling of the TP$_{\rm max}$ is shown in Fig.~4(b).
Remarkably, we find that TP$_{\rm max}$ does not decay and reaches a constant value at large $N$.
The PIM converges systematically to low-energy states even for large problems 
and delivers good approximate solutions with size-independent quality.
This property has significant implications on the time performance. 
In fact, the time-to-target is TTT$=\tau \ln(0.01)/\ln(1-{\rm TP_{\rm max}})$
where $\tau$ is the run time. 
The scaling of TTT is shown in Fig.~4(c) for $\tau=\tau_{SS}$, with $\tau_{SS}$
the mean number of iterations necessary to reach the steady state. 
TTT grows only linearly with $N$. 
Therefore, the PIM can provide near-optimal large-scale solutions within a limited time.

In conclusion, we have introduced a high-dimensional IM made by polarization oscillators.
Leveraging the dynamics of the optical field on the Poincaré sphere, 
the PIM intrinsically operates in three dimensions.
Our scheme thus reveals how to realize high-dimensional IMs by a simple photonic platform.
Simulations of the optoelectronic setup unveil a novel computing mechanism 
based on the dynamical transition of the Stokes vectors into the Ising GS.
This dimensional collapse can be extended to other optical and electromagnetic systems,
also triggering novel heuristic algorithms. The process shares aspects with quantum annealing~\cite{Nishimori1998} but it is purely classical.
Extensive numerical analysis demonstrates the PIM achieves a superior SP and an exponential improvement in scaling over CIMs as a result of its high-dimensional operation.
The performance can be further enhanced by annealing the machine parameters,
introducing noise, or by engineering the feedback signal.
Our setup can be readily implemented in experiments, where it can operate with an iteration time of nanoseconds by using available GHz electro-optic modulators.
The proposed PIM opens the way 
to a new class of combinatorial optimization hardware that exploits
the polarization of light.


\vspace*{0.1cm}
We acknowledge funding from the ERC-2024-STG LOOP  \textquotedblleft Optical polarization for ultrafast computing\textquotedblright ~No.~101162173,
HORIZON-ERC-2023-ADG HYPERSPIM No.~101139828, and EUNextGenerationEU under the
National Quantum Science and Technology Institute (NQSTI) under the National Recovery and Resilience Plan (NRRP) MUR No.~PE0000023-NQSTI, PRIN 2022597MBS PHERMIAC, and MUR PRIN 2022 MetaComputing No. 2022N738SA.

\end{document}